\documentclass[twocolumn,           
               showpacs,            
               nopreprintnumbers,     
               aps,                 
               prd,          	    
               letterpaper,             
               groupeaddress,      
               nofootinbib,         
               tightenlines,        
               floats,floatfix      
               ]{revtex4-1}

\usepackage{graphicx}
\usepackage{fancyhdr}
\usepackage[english]{babel}

\usepackage{amsmath}
\usepackage{amsfonts}
\usepackage{amssymb}
\usepackage{graphicx}
\usepackage{gensymb}
\usepackage{bm}		
\usepackage{subfigure}
\usepackage{amsfonts}
\usepackage{booktabs}
\usepackage{float} 
\newcommand{\ra}[1]{\renewcommand{\arraystretch}{#1}}

\graphicspath{ {./}{figures/}}

\begin{document}

\title{Comparison between two scalar field models using rotation curves of spiral galaxies}

\author{Lizbeth M. Fern\'andez-Hern\'andez$^{1}$}
\author{Mario A. Rodr\'iguez-Meza$^{1}$}
\email{marioalberto.rodriguez@inin.gob.mx (Corresponding author)}
\author{Tonatiuh Matos$^{2}$}
  
\address{$^1$Departamento de F\'{\i}sica, Instituto Nacional de
Investigaciones Nucleares, 
Apdo. Postal 18-1027, M\'{e}xico D.F. 11801,
M\'{e}xico}

\medskip
\address{$^2$Departamento de F\'isica, Centro de Investigaci\'on y de Estudios Avanzados del IPN, 
A.P. 14-740,
07000 M\'exico D. F., M\'exico}

\begin{abstract}
Scalar fields have been used as 
 candidates for  dark matter in the universe, from axions with masses $\sim10^{-5}$eV 
 until ultra-light scalar fields with masses $\sim10^{-22}$eV. Axions behave as cold dark matter 
 while the ultra-light scalar fields galaxies are Bose-Einstein condensate drops. The ultra-light scalar fields are also called scalar field dark matter model. 
In this work we study rotation curves for low surface brightness spiral galaxies using two scalar field models:
the Gross-Pitaevskii Bose-Einstein condensate in the Thomas-Fermi approximation 
and a scalar field solution of the Klein-Gordon equation. 
We also used the zero disk approximation galaxy model 
where photometric data is not 
considered, only the scalar field dark matter model contribution to rotation curve is taken into account.
From the best-fitting analysis of the galaxy catalog we use,
we found the range of values of the fitting parameters:
the length scale 
and the central density.
The worst fitting results (values of $\chi^2_{red}$ much greater than 1, on the average) were for the Thomas-Fermi models, 
i.e.,
the scalar field dark matter is better than the Thomas-Fermi approximation model to fit the rotation curves of the analysed galaxies.
To complete our analysis we compute 
from the fitting parameters the mass of the scalar field models and
two astrophysical quantities of interest, the dynamical dark matter mass within 300 pc and 
the characteristic central surface density of the dark matter models.
We found that the value of the central mass within 300 pc is in agreement with previous reported results, that this mass is $\approx 10^{7}$ $M_\odot/$pc$^2$, independent of the dark matter model. And, on the contrary, the value of the characteristic central surface density do depend on the dark matter model.
\end{abstract}

\date{\today}
\maketitle

\section{Introduction}\label{sec:Intro}
When 
anisotropies were discovered in the cosmic microwave background 
(CMB) 
by the Cosmic Background Explorer  (COBE) satellite 
a standard model for the universe  was almost stablished.
Planck Mission is the third generation space mission, following COBE and 
the Wilkinson Microwave Anisotropy Probe (WMAP), 
that measured the CMB anisotropies. 
Planck collaboration updated the values for the cosmological parameters, temperature and polarization anisotropies of the CMB radiation, showing that these observations are consistent  with the standard spatially-flat six-parameters model of the universe: cosmological constant 
($\Lambda$) and cold dark matter (CDM), the so called
$\Lambda \text{CDM}$ cosmology\cite{2015arXiv150201589P}.

With the study of the rotation curve (RC) of spiral galaxies \cite{Sofue:2000jx}, the existence of dark matter is gained observational support, giving rise to several models to study this  ``extra'' component of matter, not visible, that modifies the Keplerian curve, see for example \cite{Sofue:2000jx,Begeman:1991iy,Burkert:1995yz}.  
This  ``extra'' matter component known as dark matter (DM) is the difference between the mass of the galaxy 
predicted by its luminosity and its mass predicted by the rotation velocities of the stars and gas.  The study of RC gives one of the strongest evidences that spiral galaxies are embedded in extended halos of DM. 

When the evolution of the universe is studied with more detail, several models have been proposed to explain this evolution. It turns out that the most accepted model in cosmology is the $\Lambda \text{CDM}$, which uses the concept of CDM, a kind of non baryonic 
and
non relativistic particle 
at the decoupling epoch. However, it has several problems at galactic scales and less.
This is the reason why a great amount of models have arisen as alternative models to explain the DM of galaxies. One of them is the 
scalar field dark matter 
model, 
determined by a fundamental scalar field $\varphi$ \cite{2008AIPC.1083..144M}, with a minimal coupling with the metric, 
or a scalar field coupled  non minimally with the metric that have been used to study large scale structure 
formation \cite{2012AdAst2012E..29R} 
or galactic dynamics \cite{2001essf.book..213R,2007RMxFS..53d..22C,2012AIPC.1473...74R}. 
Recently, numerical simulations of large scale structure formation using a scalar field dark matter has given a strong support to these kind of dark matter models at galactic scales \cite{Schive:2014dra}. 

Scalar field based dark matter models come in different flavours depending on their equations 
of motion \cite{2008AIPC.1083..144M,2012AdAst2012E..29R,2012RMxFE..58d..53C,2011AIPC.1396..196H,2014JPhCS.545a2009M,2017arXiv170100912B}.
In this work we want to test two proposals using a fitting analysis of the rotation curves of LSB galaxies (dark matter dominated galaxies, ideal objects to test different dark matter models).
In particular,
in this paper we are going to study three scalar field based models to explain DM:
a Gross-Pitaevskii Bose-Einstein condensate in the Thomas-Fermi approximation (TF), 
with \cite{Diez-Tejedor:2014naa} and without \cite{Robles:2012uy} cut at a characteristic radial distance,  
and a scalar field solution of the Klein-Gordon equation, that
we simply refer to as scalar field dark matter model (SFDM) \cite{UrenaLopez:2002gx}.
The TF models have 
problems to fit 
the RC because the mass density profile 
may
take negative values beyond the characteristic radial distance when we do not use a cut in the density. This problem 
does not appear 
in the SFDM model 
where
the density profile is always positive.
The fitting analysis will test the models and provide the parameters that characterize the dark matter model density profiles, one of them is the mass of the scalar field.
 In order to do so, this work is organized as follows. In section
\ref{sec:STT} we briefly explain the 
two
scalar field models and the  different approaches
to be compared, the Thomas-Fermi approximation and the scalar field dark matter. 
In section \ref{sec:Results} we 
make the comparison
between the 
TF and SFDM
models 
and we find the best fitting 
parameters 
using a 
LSB galaxy
RC catalog.
In section \ref{sec:Conclusions} we
give 
our
conclusions of this comparison.


\section{The scalar field models}\label{sec:STT}

For the  SFDM model we use
the density profile  as given by \cite{UrenaLopez:2002gx}
\begin{equation}
\label{density-sfdm}
\rho_{SFDM} (r)=\rho_s \frac{\sin^2(r/r_s)}{(r/r_s)^2}\, .
\end{equation}
known as the flat oscillaton.
This corresponds to the static solution of 
the corresponding Klein-Gordon equation
with a quadratic potential for the scalar field $\varphi$, $V \left(\varphi\right) = m_\varphi^2 \varphi^2/2$, $m_\varphi$ is the scalar field mass and is given by the inverse of $r_s$ through the Compton relationship $r_s = \hbar c/m_\varphi$, $\hbar$ is the Planck's constant and $c$ the speed of light. 
See
Ref. \cite{UrenaLopez:2002gx} for details.

The 
corresponding
rotation velocity 
is
\begin{equation}
\label{velocity-sfdm}
V_{SFDM}^2 (r)=-\frac{\rho_s}{4} r_s^2 \left(-2 + \frac{r_s}{r}\frac{\sin(2r/r_s)}{r/r_s}\right)\, .
\end{equation}

The other SF model we want to consider is the scalar field that satisfies the Gross-Pitaevskii and Poisson equations. These two equations have, 
in the Thomas-Fermi approximation,
the solution \cite{Harko:2011xw}
\begin{equation}
\label{TF-approx}
\rho_{TF}(r) = \rho_s \frac{\sin(kr)}{kr}\, ,
\end{equation}
where $\rho_s$ is the central density of the condensate, $\rho_s = \rho(0)$. The total mass of the galactic dark matter halo is
\begin{equation}
\label{TF-mass}
M_{TF}(r) = \frac{4\pi \rho r}{k^2}\left(1-kr \cot (kr)\right)\, .
\end{equation}
This expression has a relevant significance only when the mass distribution is inside of a boundary surface with radius $R$  (Thomas-Fermi approximation with a cut), where
$\rho(R)=0$ and then $kR=\pi$
\begin{equation}
\label{TF-massSF}
R=\pi\sqrt{\frac{\hbar^2 a}{Gm_\psi^3}}\, ,
\end{equation}
where $G$ is the Newton gravitational constant and
$a$ and $m_\psi$ are the scattering length and the mass of the scalar field particle, respectively.
This expression fix the radius of the condensate dark matter halo. We follow  \cite{Harko:2011xw} and chose $a=1$ fm.
Hence, the tangential velocity is
\begin{equation}
\label{TF-velocity}
V_{TF}^2(r)=\frac{4G\rho_s R^2}{\pi}\left[\frac{\sin{(\pi r/R)}}{\pi r/R}  - 
\cos\left({\frac{\pi r}{R}}\right)\right]\, ,
\end{equation}
for $r>R$ the rotation curves follow the standard Keplerian law.

If we considered that the previous results are valid even outside of the boundary surface 
(Thomas-Fermi approximation without a cut \cite{Robles:2012uy}),  we have two scalar field models in the TF approximation. In this case we will have 
that equation
(\ref{TF-velocity}) is without constriction.

Therefore, for spherical symmetry, 
our SF galaxy model will be as follows. A test particle will move under the action of the 
potential $\Phi_{DM} (r)$  which is the potential due to a DM mass distribution,  where
\begin{equation}
\label{theoretical-potentials}
\Phi_{DM} (r) = 
\left \{ \begin{matrix} 
 \Phi_{TF} &= \left \{ \begin{matrix} 
\text{with cut} &\\
\text{without cut} &
\end{matrix}\right.
\\
\\ \Phi_{SFDM} & 
\end{matrix}\right.  
\end{equation}

The potential of the SF, $\Phi_{DM} (r)$ is the solution the corresponding Klein-Gordon equation for the SFDM 
model \cite{UrenaLopez:2002gx} or it is the solution for Gross-Pitaevskii and Poisson equations in the Thomas-Fermi approximation \cite{Harko:2011xw}.

The rotational velocities as given in Eqs. (\ref{velocity-sfdm}) and (\ref{TF-velocity}) (with and without a cut) were obtained using  
\begin{equation}
\label{rotation-curve}
V^2 = r\bigg{|}\frac{d\Phi_{DM}}{dr}\bigg{|}\, .
\end{equation}



\section{Results}\label{sec:Results}

In this section we perform the comparison among models in Eq. (\ref{theoretical-potentials})
that give the theoretical RC that we test against observed RC by a fitting analysis.

We are using a high resolution LSB galaxies catalog reported in \cite{deBlok:2001mf}. 
LSB galaxies are the ideal objects to test the distribution of dark matter in galaxies and to compare different
dark matter models.
The authors in Ref. \cite{deBlok:2001mf}  considered the visible data contribution
for classifying the galaxies 
according to availability of
photometric data 
and use three models: zero disk model, 
where the dark matter is the 
main
component in the
halos, making zero all the visible components of the galaxies; 
galaxies with photometry with a constant ratio $M/L_*$ and the maximum disk model.
In our case only galaxies without photometry data are considered \cite{KuziodeNaray:2007qi} and we will use the zero disk model.   

Therefore in this work we  use only the 
zero disk
model where the velocity contributions from baryons are ignored, they are test particles, 
and the total RC has only the contribution of the dark matter models.
Accordingly we consider only galaxies classified as without photometric data in Ref. \cite{deBlok:2001mf}.
These galaxies belong to the sample for which an optical rotation curve was possible to measure 
but there were no optical nor H I photometry, implying that mass models for the stars or gas were not 
possible to build.

We perform a minimum $\chi^2$ analysis
to fit the observations with $\chi^2 = \sum_{i=1}\left(
\frac{V_i - V(r_i)}{\sigma_i}
\right)^2\, ,$
where $V_i$ is the observed  velocity, $\sigma_i $ is the error in the measurement  of $V_i$
and $V(r_i)$ is the theoretical value of the velocity, 
which is computed at the same position
where $V_i$ was measured. For $V(r_i)$ we will use each one of the models in Eq. (\ref{theoretical-potentials}).
We use units where the length is in kpc, the velocity in km/s and $G=1$. 
The free parameters to fit are the following:
for  the SFDM model,
the RC equation is given by Eq. (\ref{velocity-sfdm}), it has two parameters, $\rho_s$ and $r_s$;
for the TF models the RC is given by Eq. (\ref{TF-velocity}) then it has also two parameters, $\rho_s$ and $R$. 
In the fitting analysis we impose that all these parameters are always positive and without an upper limit.
To complete the data analysis we compute the errors in the fitting parameters and the $Q$-value that gives us the goodness of fit \cite{numrecip}.

 \begin{figure*} 
    \begin{center}$
    \begin{array}{ccc}
\includegraphics[width=1.95in]{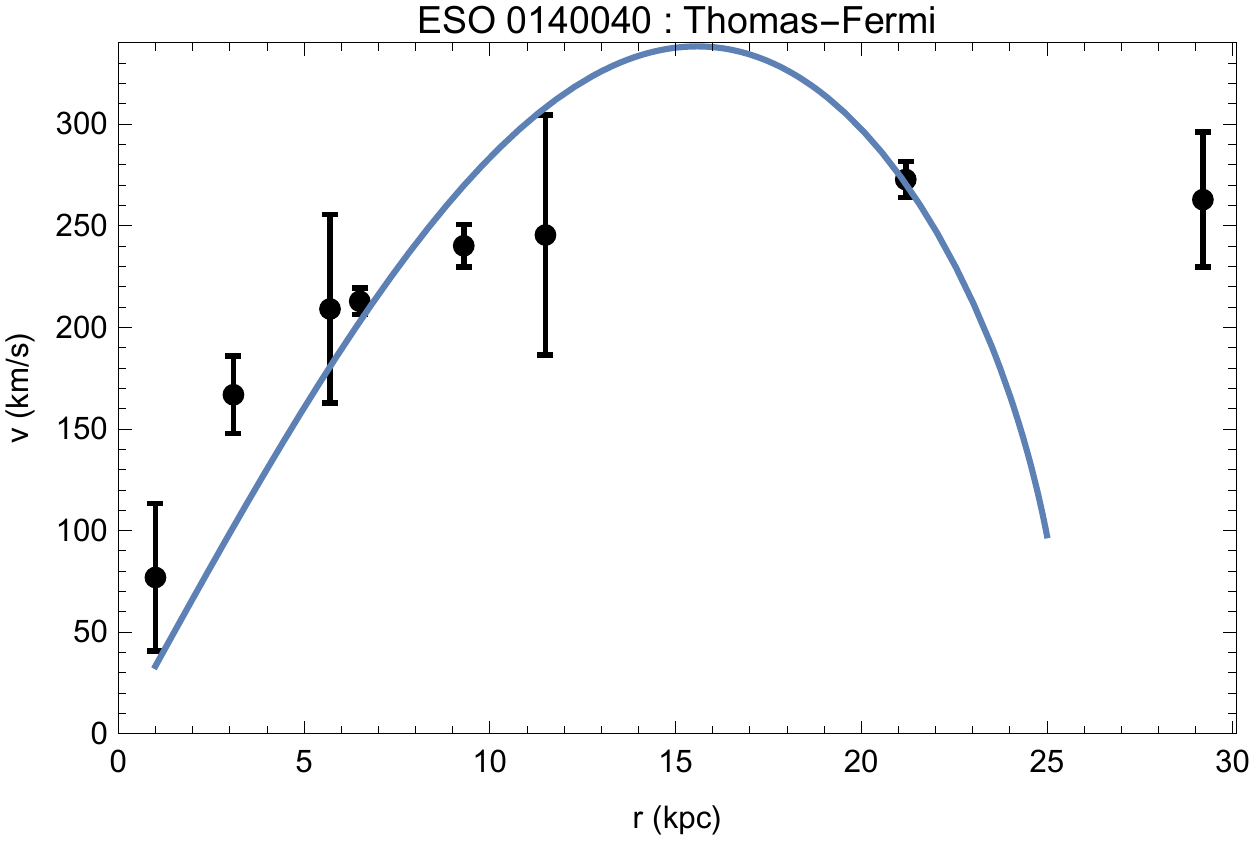} &
\includegraphics[width=1.95in]{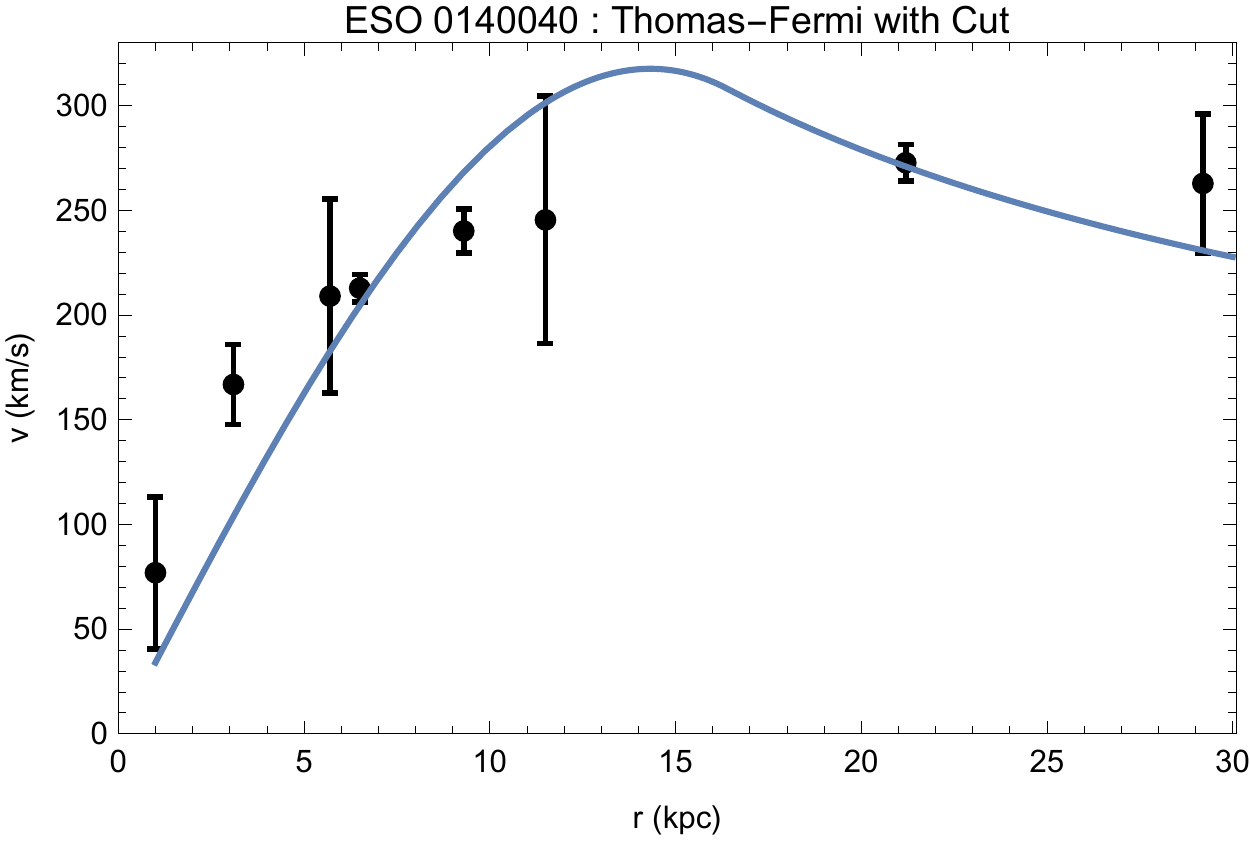} &
\includegraphics[width=1.95in]{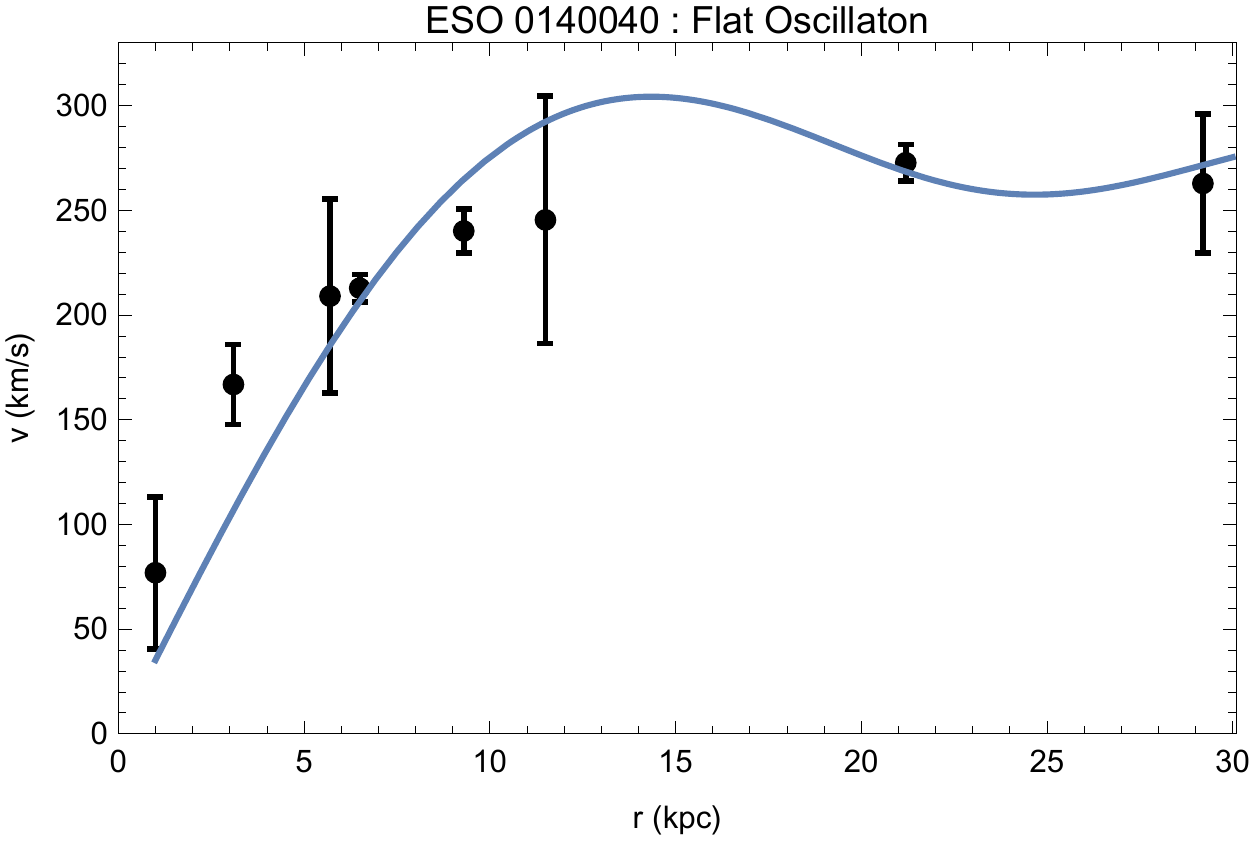} \\
\includegraphics[width=1.95in]{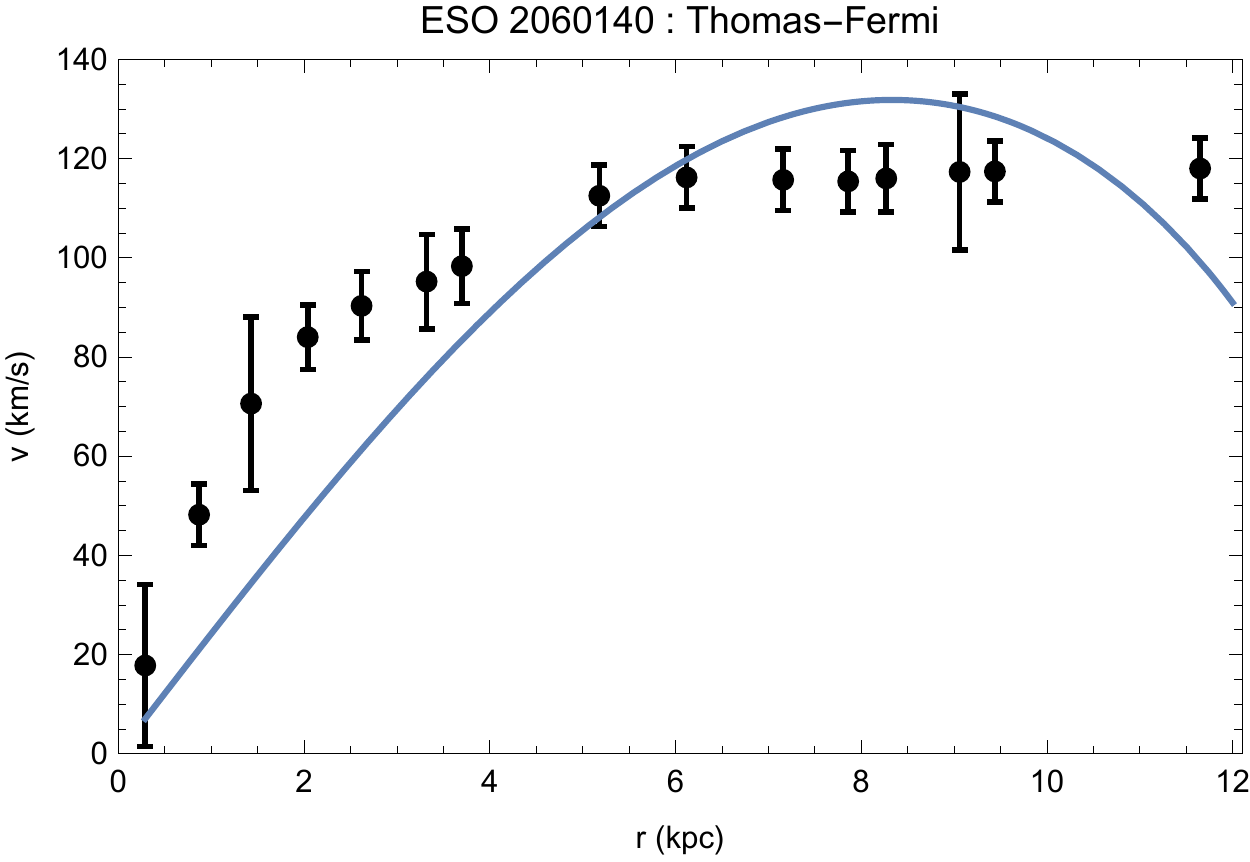} &
\includegraphics[width=1.95in]{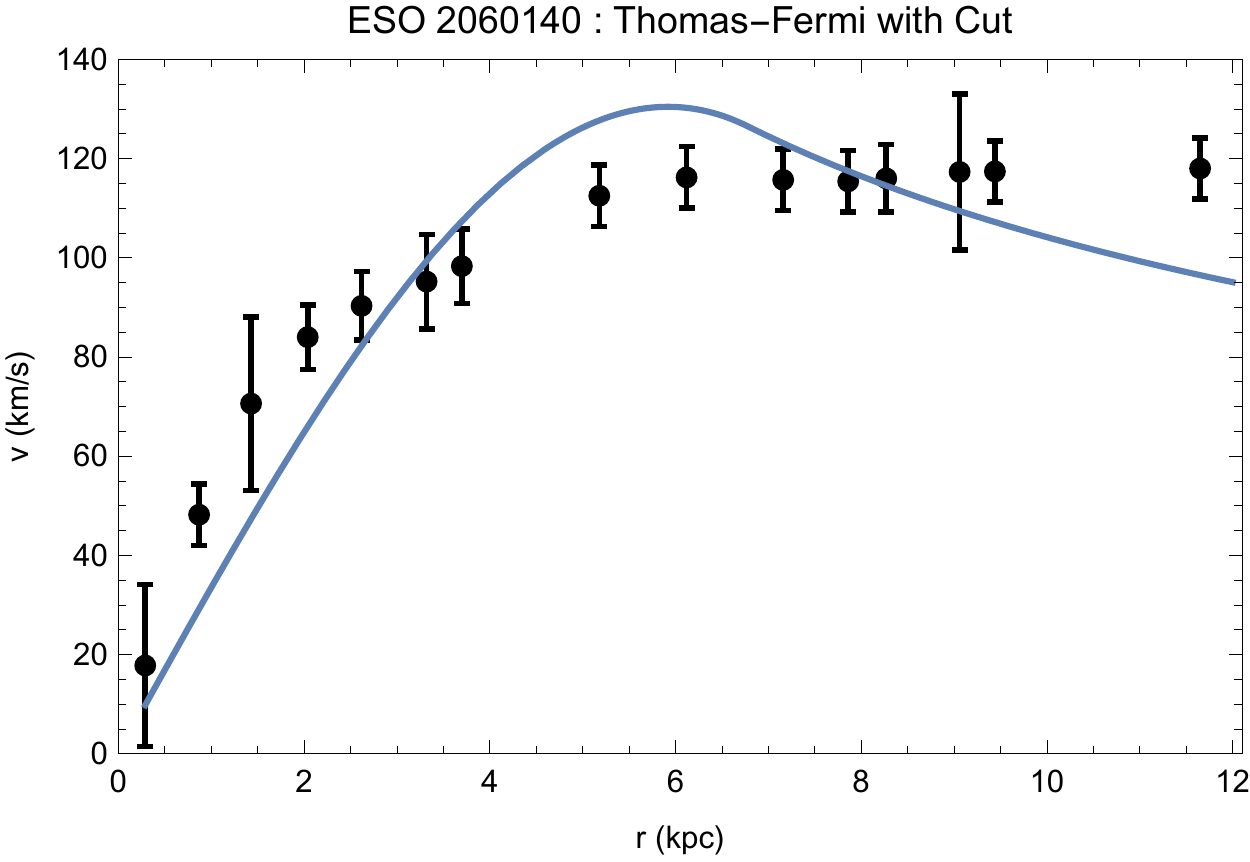} &
\includegraphics[width=1.95in]{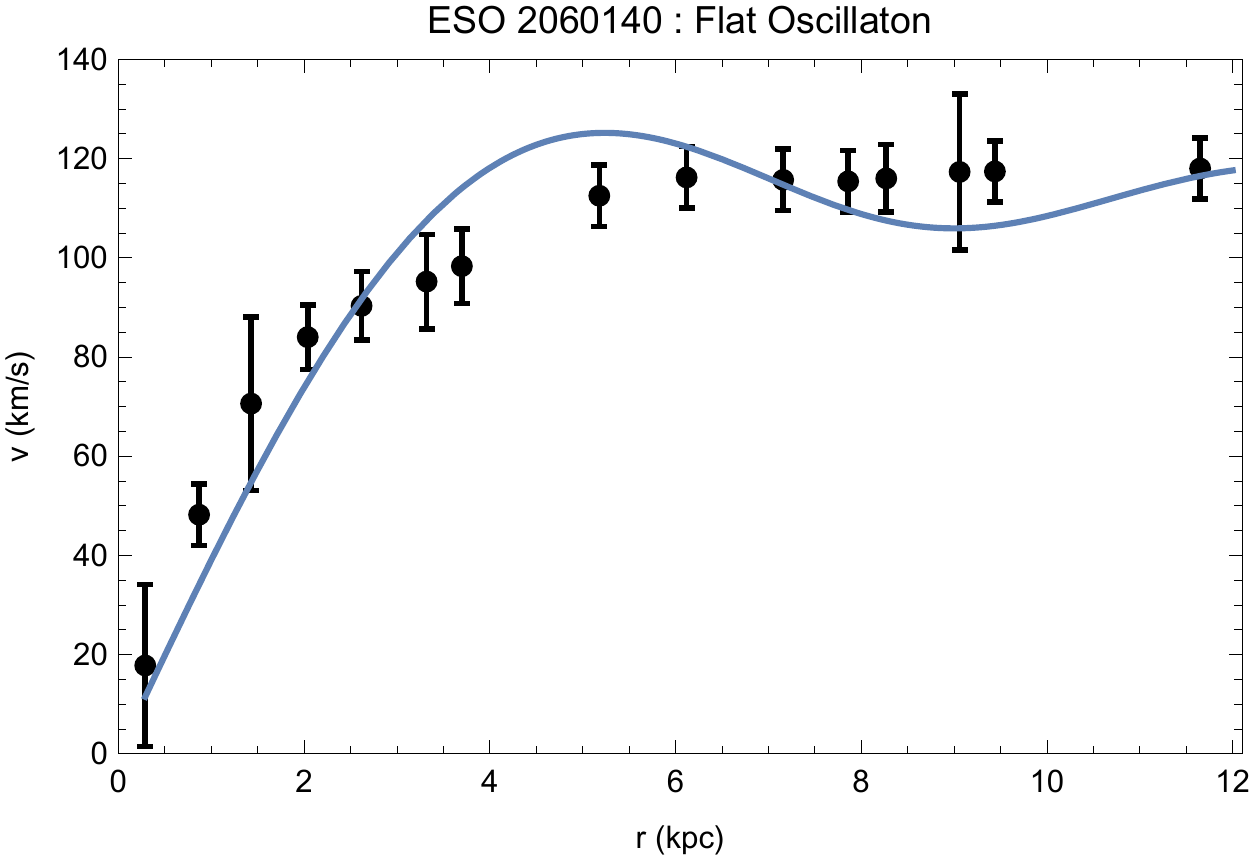} \\
\includegraphics[width=1.95in]{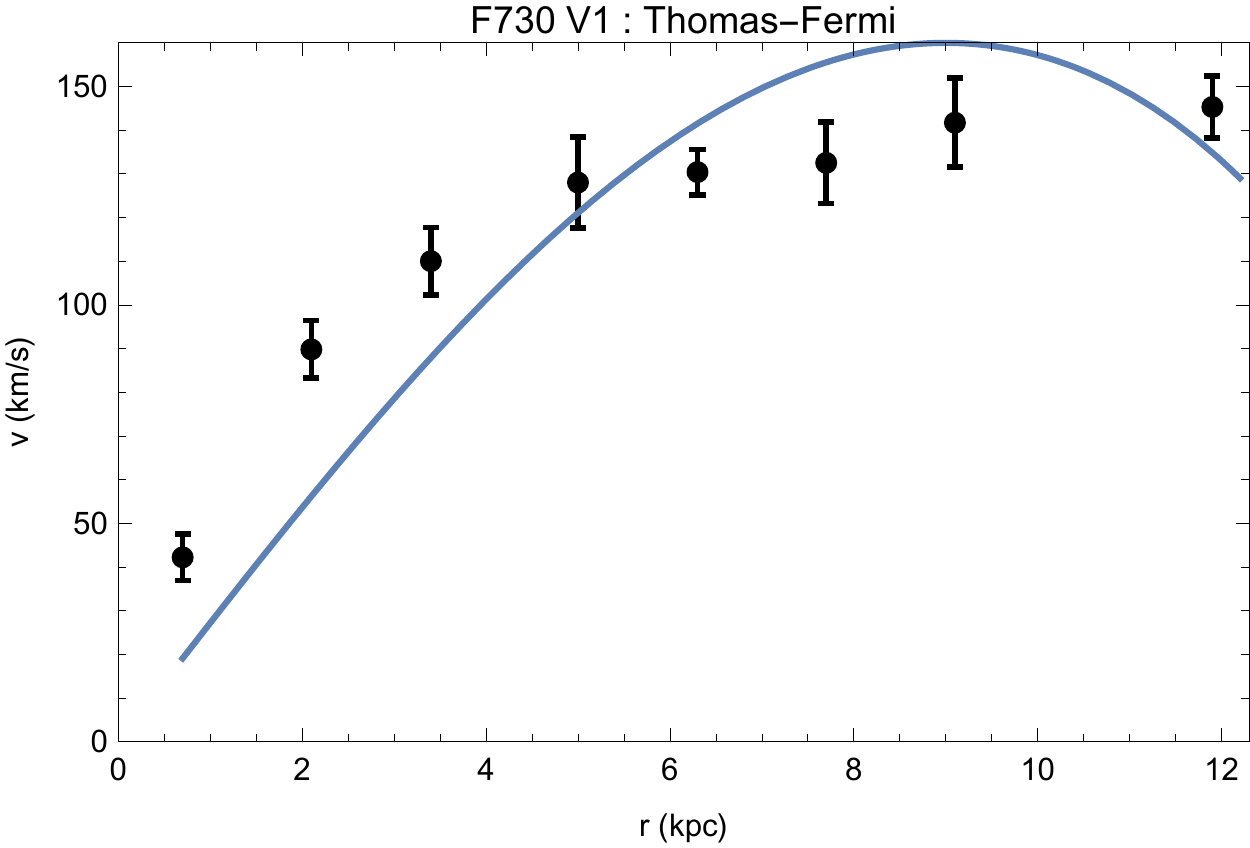} &
\includegraphics[width=1.95in]{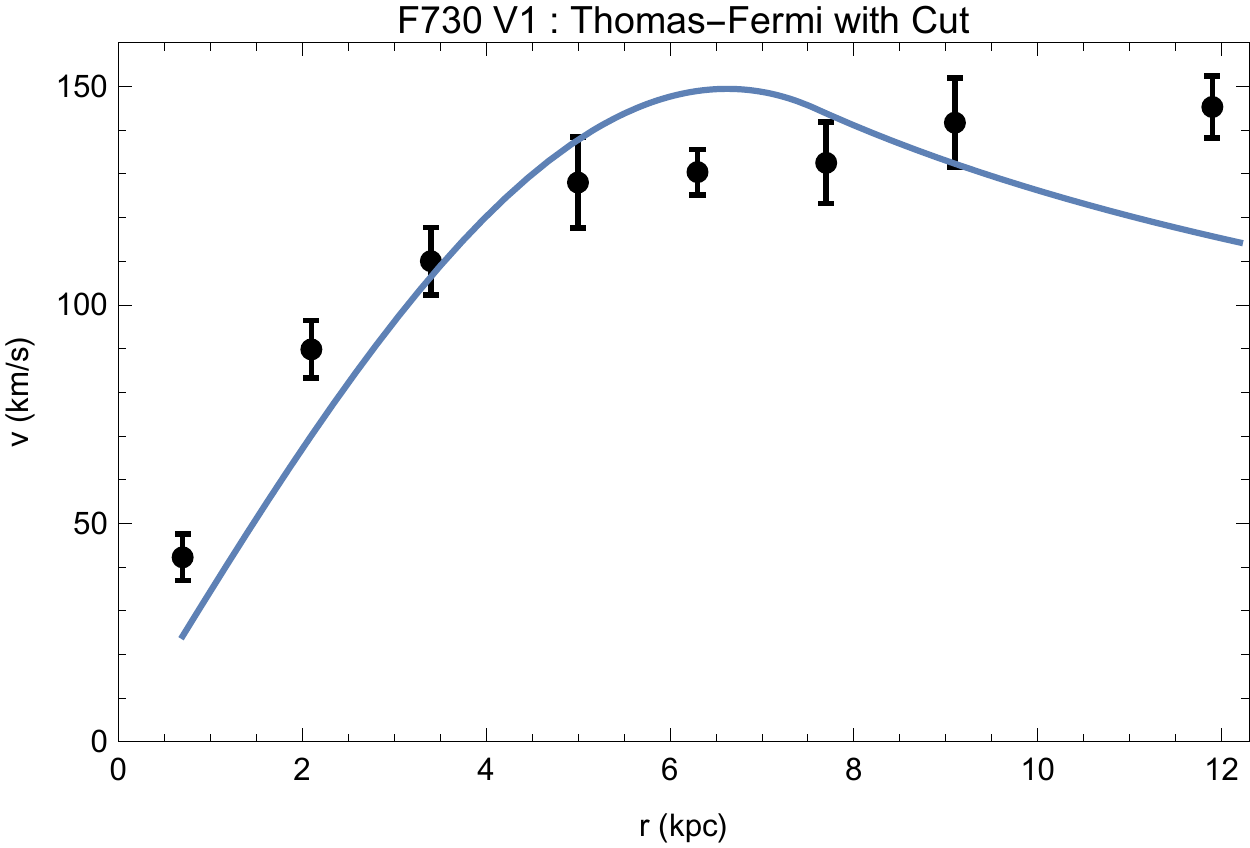} &
\includegraphics[width=1.95in]{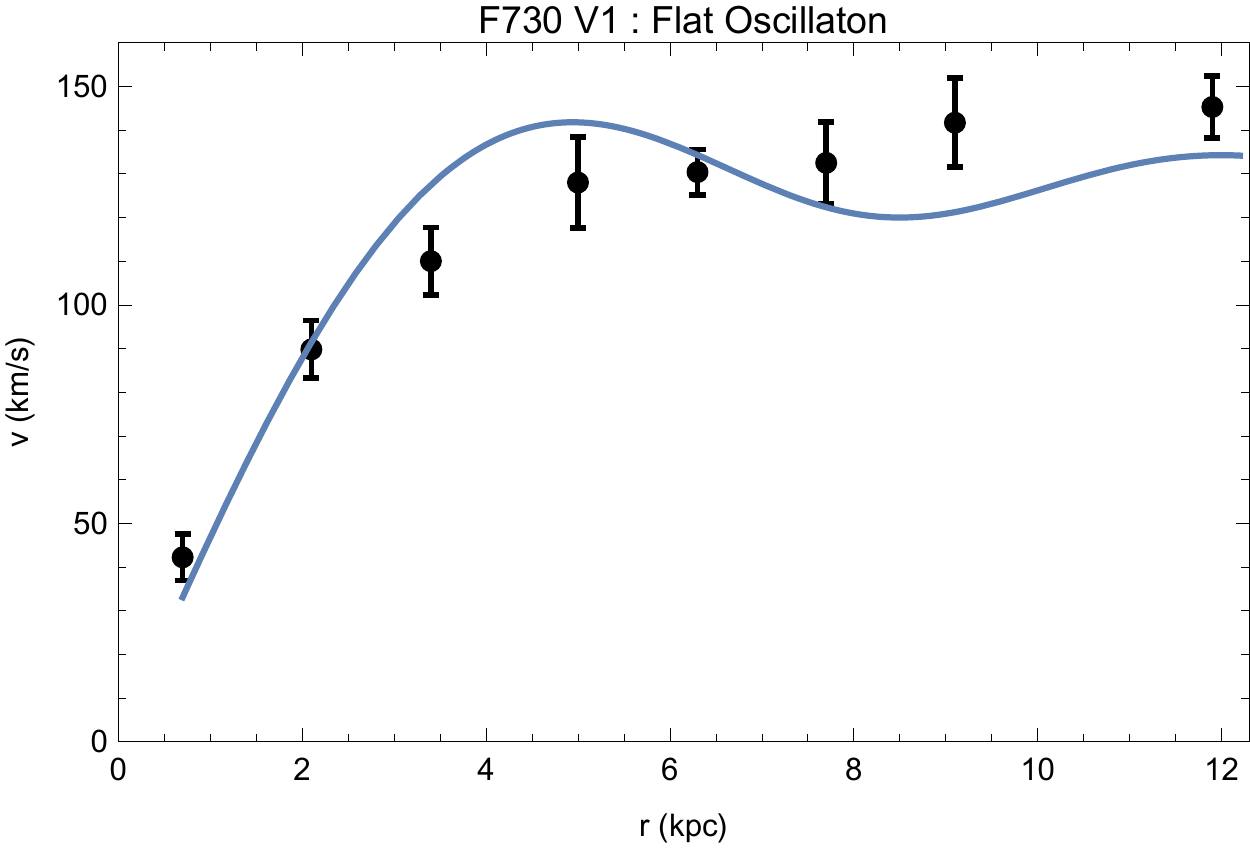} \\
\includegraphics[width=1.95in]{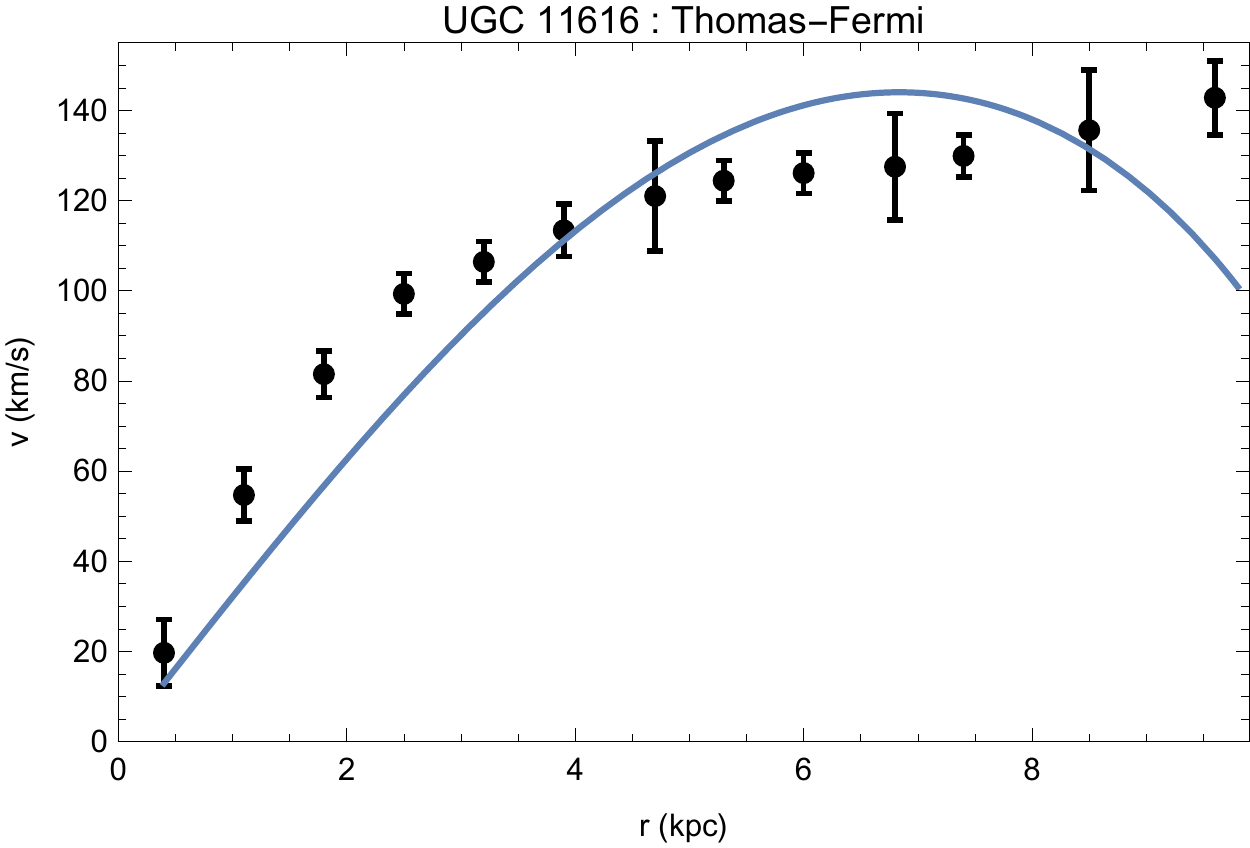} &
\includegraphics[width=1.95in]{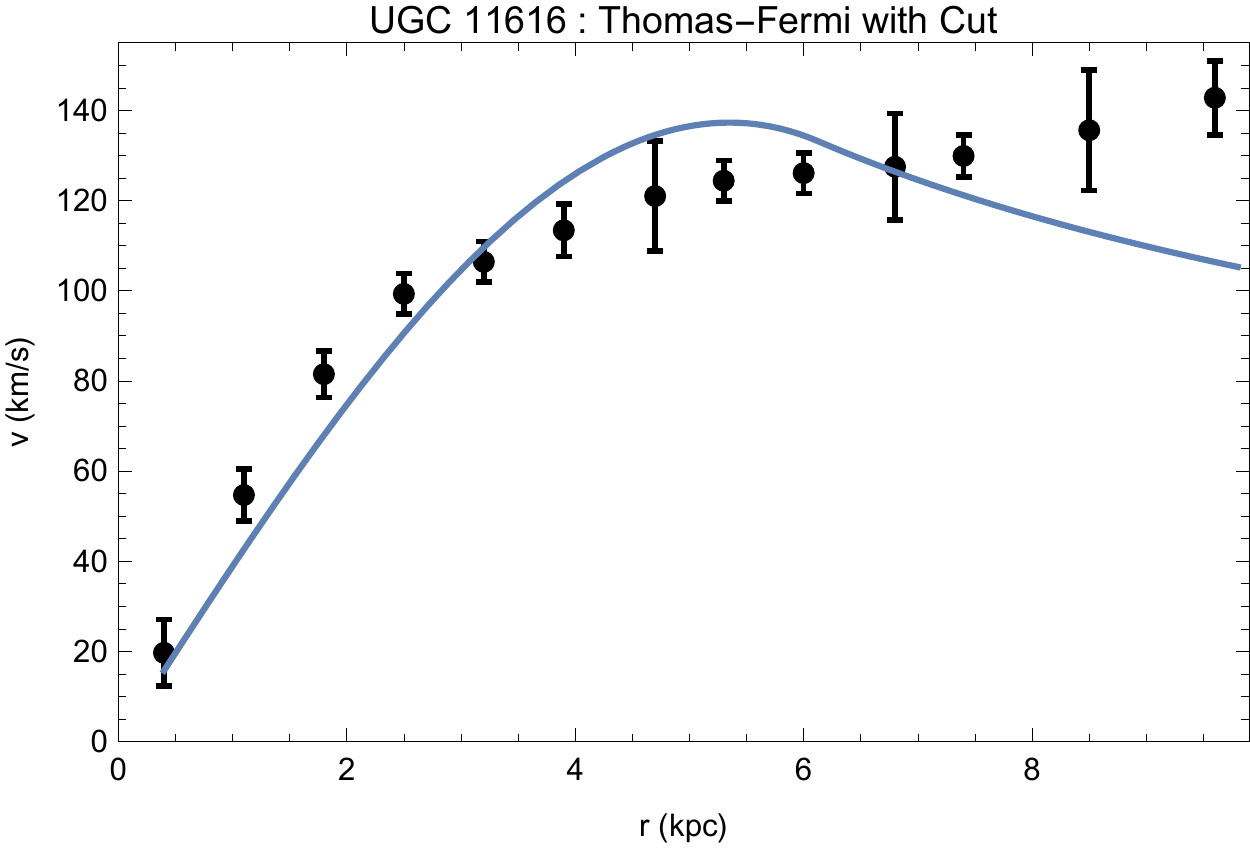} &
\includegraphics[width=1.95in]{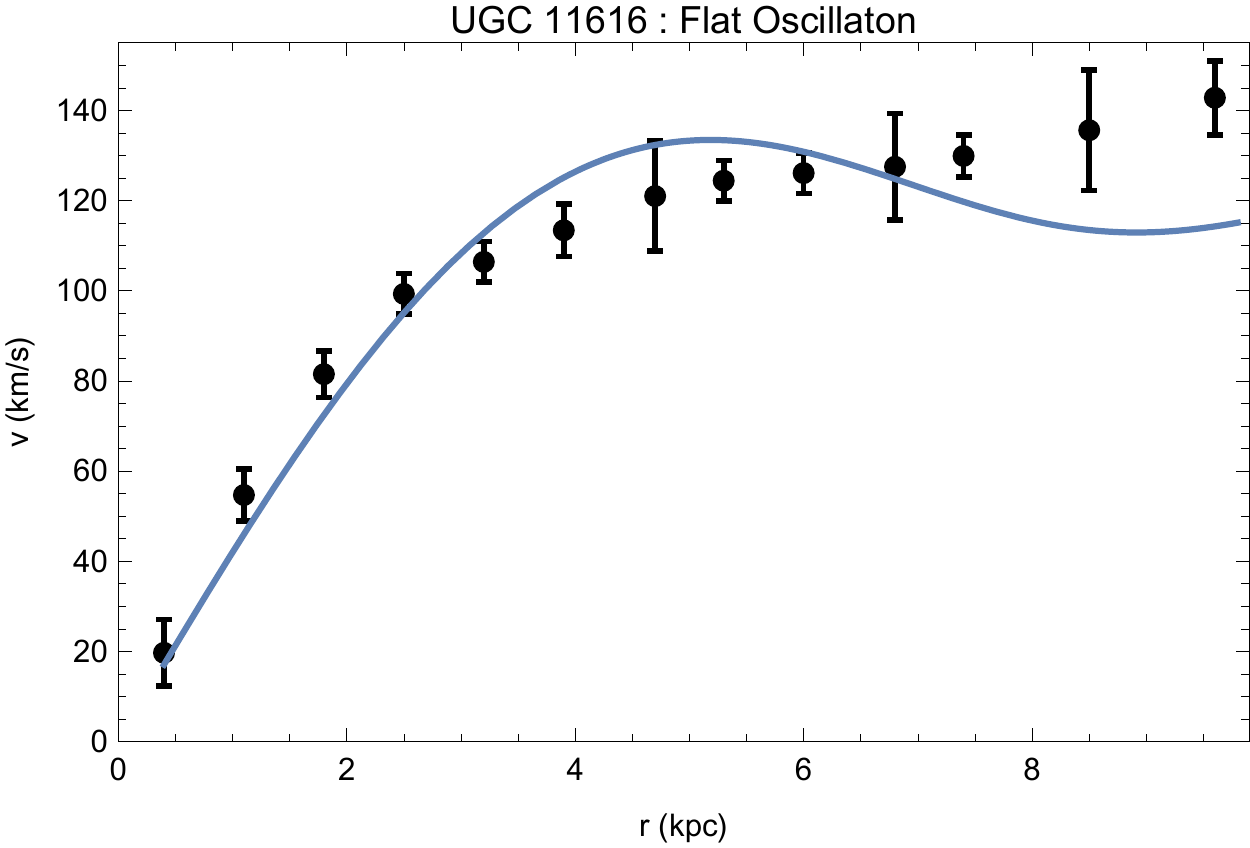} \\
\includegraphics[width=1.95in]{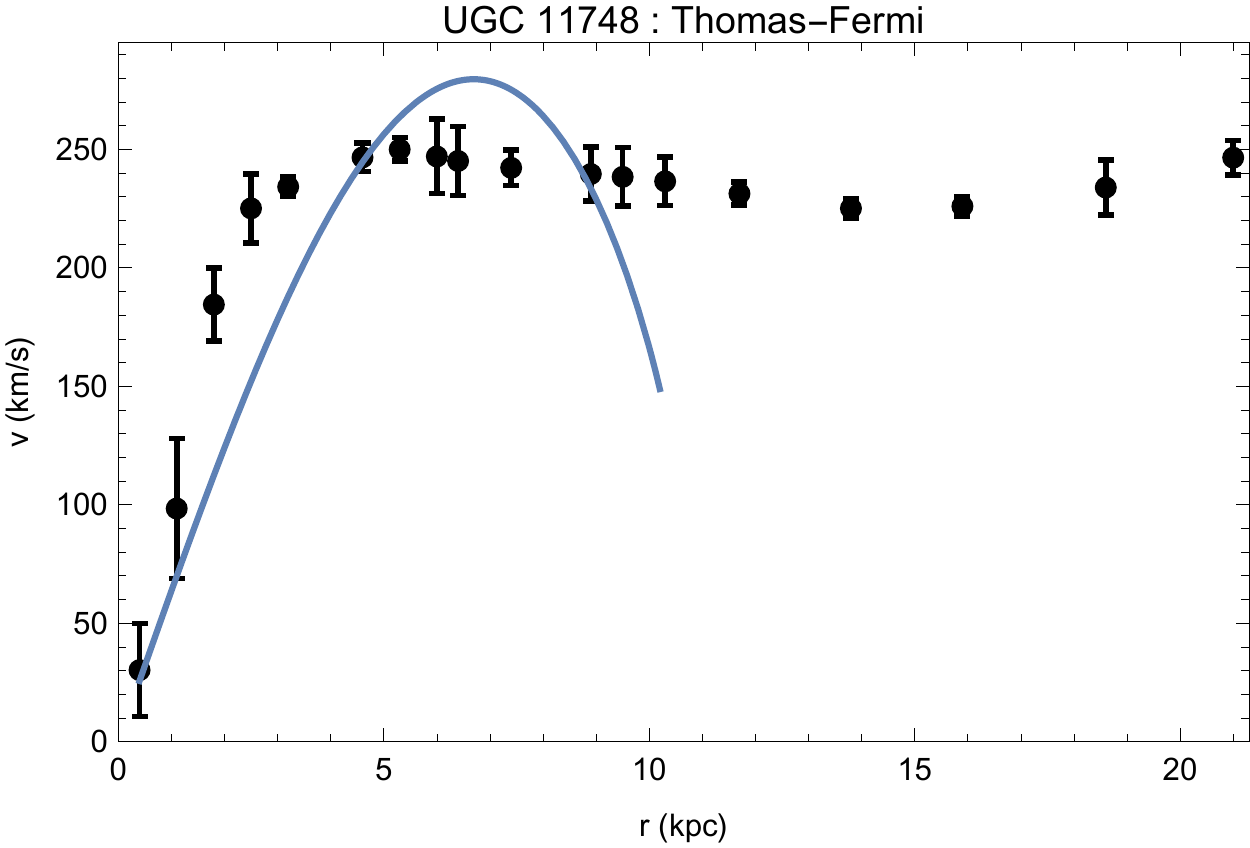} &
\includegraphics[width=1.95in]{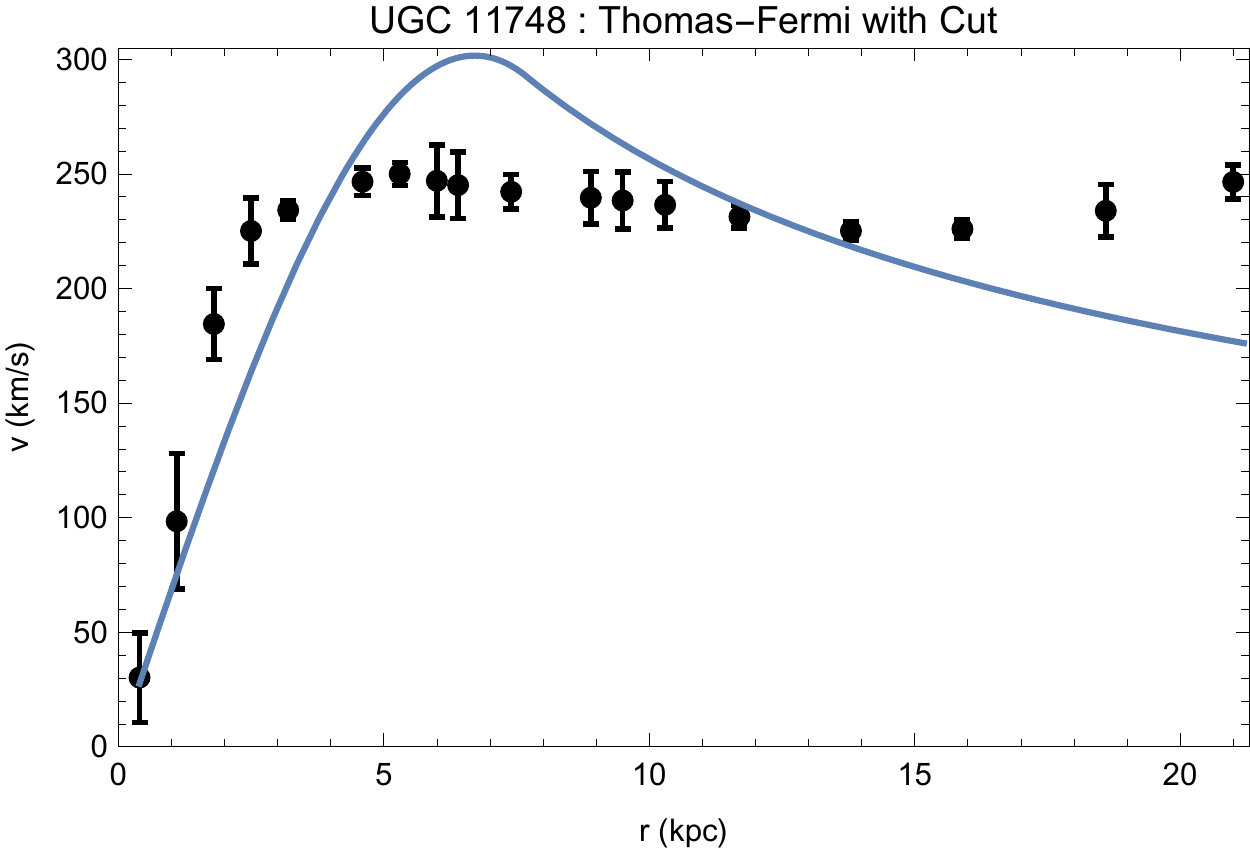} &
\includegraphics[width=1.95in]{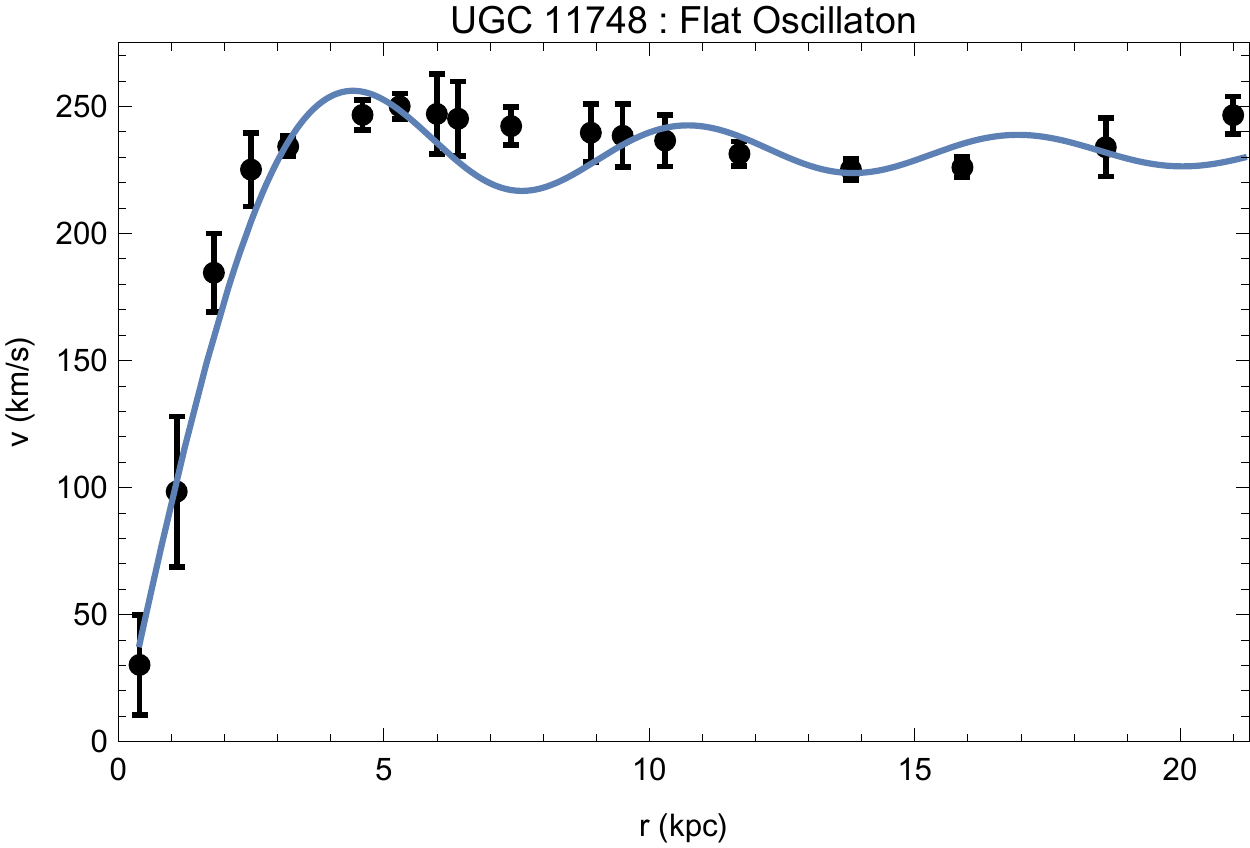} \\
\includegraphics[width=1.95in]{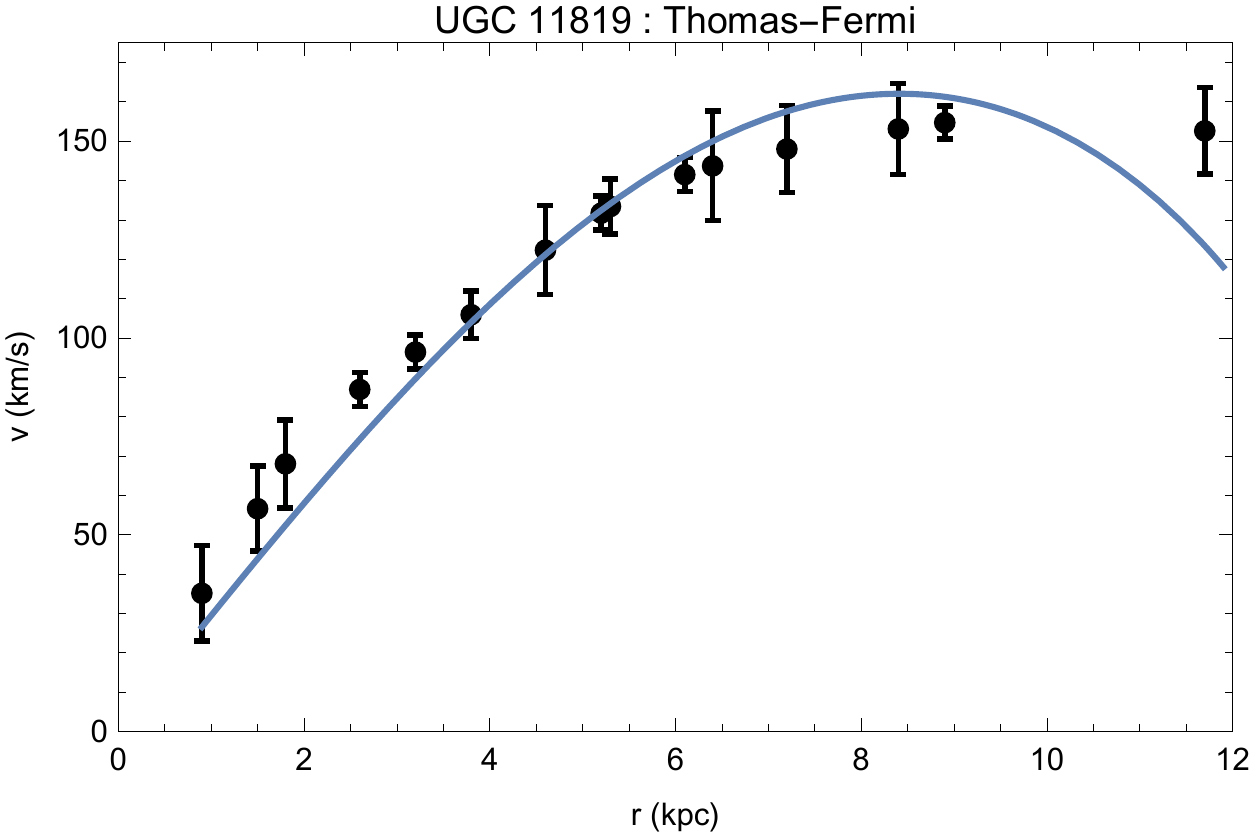} &
\includegraphics[width=1.95in]{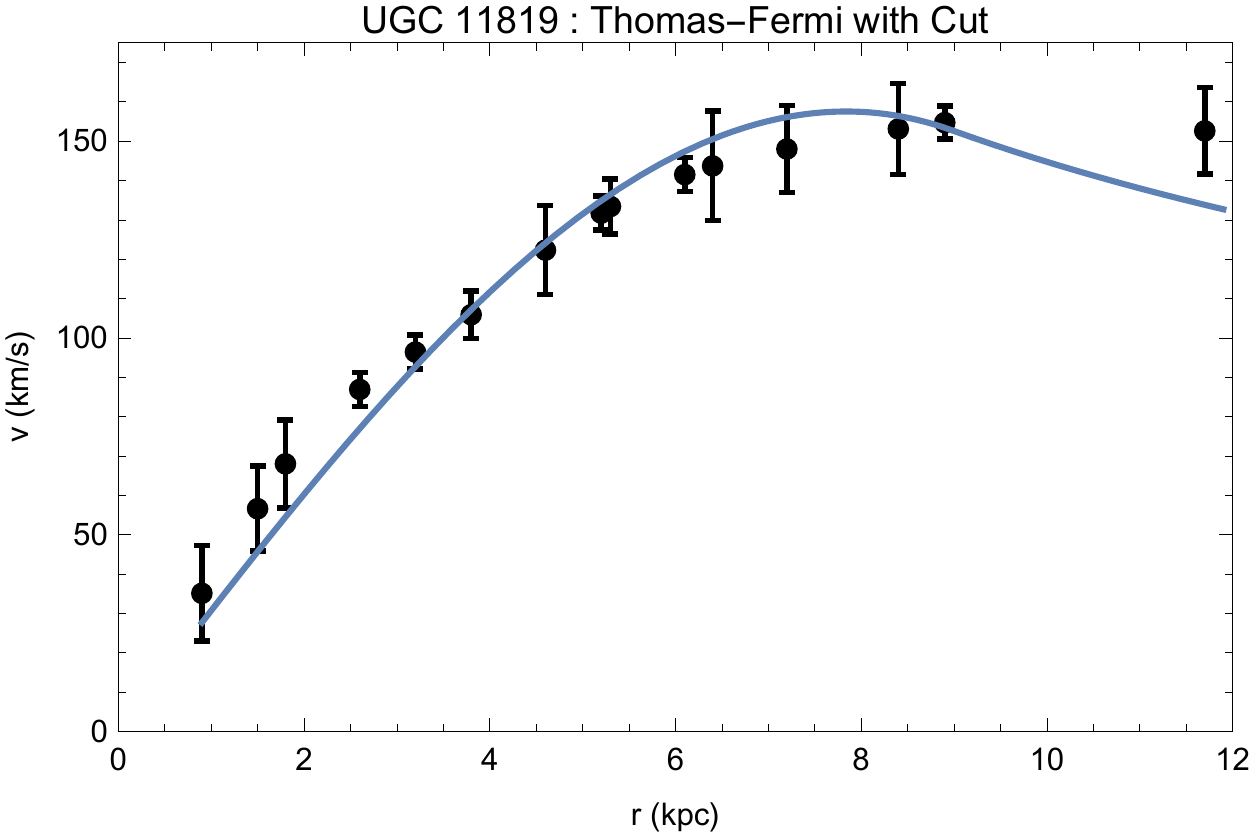} &
\includegraphics[width=1.95in]{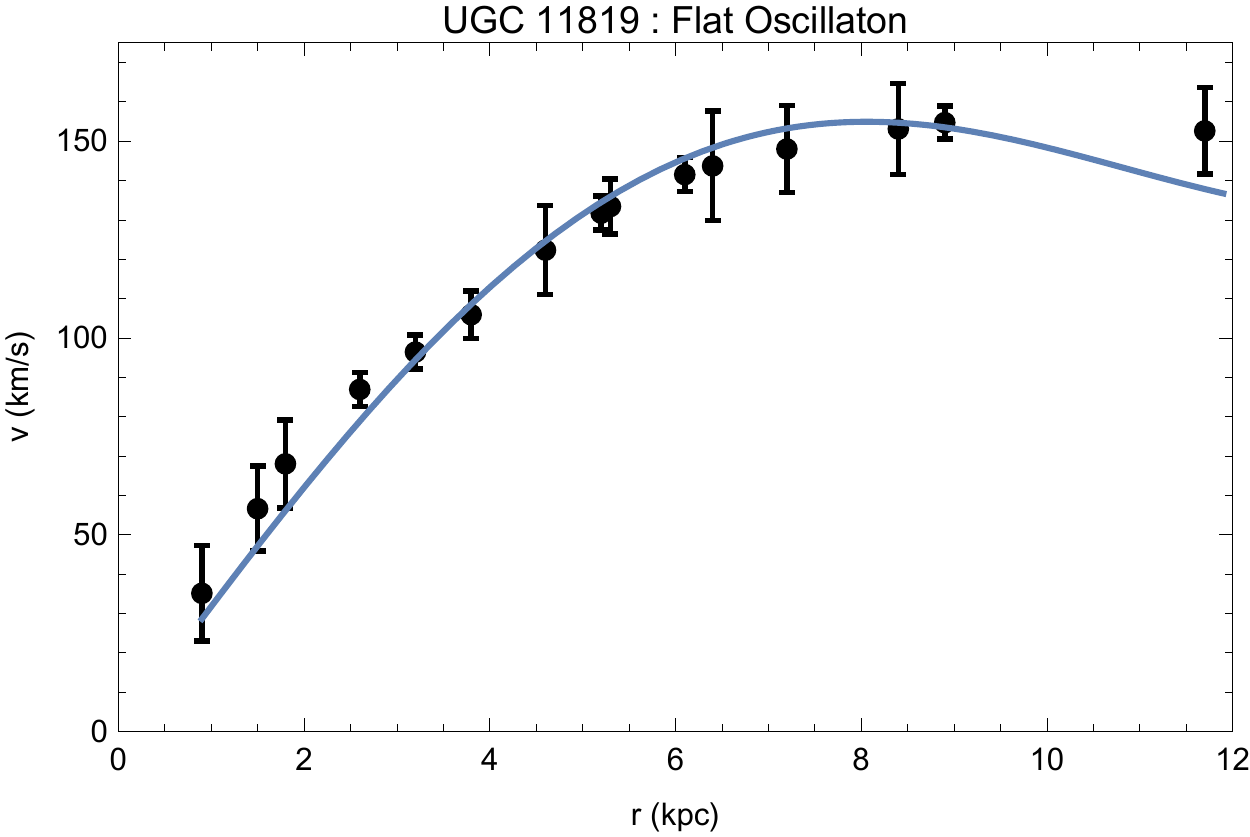} 
 \end{array}$
    \end{center}
   \caption{LSB galaxies fit with the zero disk and the scalar field models, Eq. (\ref{theoretical-potentials}). Black solid circles are the observed RC. The blue line is the theoretical RC obtained by the fitting method.}
    \label{fig:SFDM-md}
    \end{figure*}

In Fig. (\ref{fig:SFDM-md}) we show the comparison between the observed RC and the theoretical RC as given by fitting each of the DM models for the galaxies without photometry. The left column show the TF model without cut, the middle column the TF model with cut and the right column SFDM flat oscillaton model. It is very clear that the TF approximation without cut is the worst case, as can be seen, for instance, for galaxies ESO 0140040 and UGC 11748 whereas the best fitting results were obtained for flat oscillaton model as can be see in the column on the right. This is confirmed in Tables \ref{tab:TF-zerodisk}--\ref{tab:SFDM-zerodisk} (see below) in columns (9) where are shown the $Q$-values, for the TF without cut model they are very small or zero for a lot of galaxies.

Table \ref{tab:TF-zerodisk} shows the best fitting parameters for galaxies without photometry 
for the scalar field dark matter model in the TF approximation without cut, using Eq. (\ref{TF-velocity}).  
Fitting parameters are shown in columns (2) and (3). From column (3) and using Eq. (\ref{TF-massSF}) (we have chosen $a=1$ fm according to Ref. \cite{Harko:2011xw}) we obtain column (4), the mass of the scalar field particle, its values are in the range of $9.86$ meV $< m_\psi < 51.11$ meV, consistent with the upper limit $m_\psi< 1.87$ eV computed from cosmological considerations \cite{2008PhRvD..77d3518B}. Fitting parameters are consistent with values found in Ref.  \cite{Robles:2012uy} except for galaxies, UGC 11616, UGC 11748 and UGC 11819, due to that in Ref.  \cite{Robles:2012uy}  some measured values are missing. This DM model has the problem that density can take negative values and so the rotation velocity (leading to instabilities) as can be noticed clearly in Fig.  (\ref{fig:SFDM-md}) left column, galaxies ESO 0140040 and UGC 11748.
In column (5) we show the characteristic surface density, $\mu_{DM}=\rho_s r_c$, $r_c$ is the radius at which the density 
has decayed to a half of the central density value, $\rho_s$. Its mean value is, $225.7$ $M_\odot/$pc$^2$.
In column (6) it is shown the dynamical mass of the dark matter up to 300 pc. Its mean value is $0.6\times 10^7$ $M_\odot$ in agreement with the value found for dSph Milky Way satellites \cite{2008Natur.454.1096S}. And in column (7) we show the total mass of this DM model.

We make the same analysis for Table \ref{tab:TF-zerodisk-cut} which present the results for the TF approximation with a cut. The scalar field mass is in the range, $10.43$ meV $< m_\psi < 51.11$ meV, almost the same as the TF without a cut. The mean value of the characteristic surface density is $244.4$ $M_\odot/$pc$^2$ and the mean value of the mass within 300 pc is $0.7\times 10^7$ $M_\odot$.

For 
the SFDM model we can compute the
same statistical information as was presented  above, the results are shown in Table \ref{tab:SFDM-zerodisk}.
In this case the characteristic surface density is $\mu_{DM}=\rho_s r_s$, $r_s$ is the scale radius and $\rho_s$ is the central density value of the flat oscillaton. Its mean value is $186.7$ $M_\odot/$pc$^2$. The DM mass within 300 pc mean value is $0.9\times 10^7$ $M_\odot$ that is also in agreement with the result found in Ref. \cite{2008Natur.454.1096S}.
The range of the oscillaton mass is $1\times 10^{-27}$ eV $< m_\varphi < 10.45\times 10^{-27}$ eV. The mass of the flat oscillaton was computed using the Compton wave length relationship~\cite{UrenaLopez:2002gx}, $r_s$ shown in column (3) of Table  \ref{tab:SFDM-zerodisk}. This range of values for the flat oscillaton rule out this model according to the results presented in Refs. \cite{2012RMxFE..58d..53C,2017arXiv170100912B}.

In column (7) of  Tables \ref{tab:TF-zerodisk}--\ref{tab:SFDM-zerodisk} we show the estimated values of the DM models total mass. Their mean values are: for the TF model without cut, $6.4\times 10^{10}$ $M_\odot$, for the TF model with a cut, $4.7\times 10^{10}$ $M_\odot$ and $6.5\times 10^{10}$ $M_\odot$ for the oscillaton model. All these value are very similar and are consistent for the values of LSB galaxies.

\begin{center}
\begin{table*}
\ra{1.3}
\begin{center}
\resizebox{16cm}{!} {
\begin{tabular}{@{}l r r r r r r r r r r @ {}}\toprule
    \hline \hline
    \multicolumn{9}{c}{Galaxy sample} \\
    \multicolumn{9}{c}{Thomas-Fermi DM Model}\\
    \hline 
   \multicolumn{9}{c}{Best Fitting Parameters}\\
   \hline 
\multicolumn{1}{c}{Galaxy} &
\multicolumn{1}{c}{$\rho_{s}$ }&
\multicolumn{1}{c}{$R$ } &
\multicolumn{1}{c}{$m_{\psi}$} &
\multicolumn{1}{c}{$\mu_{DM}$} &
\multicolumn{1}{c}{$M_{DM}(300$ pc$)$}& 
\multicolumn{1}{c}{$M_{DM}(r_{\text{max}})$}& 
\multicolumn{1}{c}{$\chi^2_{\text{red}}$} &
\multicolumn{1}{c}{$Q$-value}\\
\multicolumn{1}{c}{ }& 
\multicolumn{1}{c}{($10^{-3}M_\odot/$pc$^3$)}&
\multicolumn{1}{c}{(kpc) }& 
\multicolumn{1}{c}{(meV)}& 
\multicolumn{1}{c}{($M_\odot/$pc$^2$) }& 
\multicolumn{1}{c}{($10^{7}M_\odot$)}& 
\multicolumn{1}{c}{($10^{11}M_\odot$)}& 
\multicolumn{1}{c}{ }\\
\multicolumn{1}{c}{(1)}& 
\multicolumn{1}{c}{(2)}& 
\multicolumn{1}{c}{(3)}& 
\multicolumn{1}{c}{(4)}& 
\multicolumn{1}{c}{(5)}& 
\multicolumn{1}{c}{(6)}& 
\multicolumn{1}{c}{(7)}& 
\multicolumn{1}{c}{(8)}& 
\multicolumn{1}{c}{(9)}\\
 
\cmidrule[0.4pt](r{0.25em}){1-1}
\cmidrule[0.4pt](lr{0.25em}){2-2}
\cmidrule[0.4pt](lr{0.25em}){3-3}
\cmidrule[0.4pt](lr{0.25em}){4-4}
\cmidrule[0.4pt](lr{0.25em}){5-5}
\cmidrule[0.4pt](lr{0.25em}){6-6}
\cmidrule[0.4pt](lr{0.25em}){7-7}
\cmidrule[0.4pt](lr{0.25em}){8-8}
\cmidrule[0.4pt](lr{0.25em}){9-9}

 ESO 0140040 & 61.71$\pm$4.98 & 17.84$\pm$0.37 & 9.86 & 664.20 & 0.70 & 5.25 & 4.15 & 0.0004 \\ 
 ESO 0840411 & 4.69$\pm$1.35 & 11.82$\pm$3.61 & 12.98 & 33.43 & 0.05 & 0.08 & 0.10 & 0.9982 \\ 
 ESO 1200211 & 13.66$\pm$3.84 & 2.92$\pm$0.41 & 32.95 & 24.07 & 0.15 & 0.003   & 0.44 & 0.9476 \\ 
 ESO 1870510 & 32.55$\pm$7.35 & 2.93$\pm$0.38 & 32.85 & 57.63 & 0.36 & 0.01 & 0.21 & 0.9932 \\ 
 ESO 2060140 & 32.78$\pm$2.67 & 9.54$\pm$0.29 & 14.97 & 188.69 & 0.37 & 0.23 & 8.43 & 0.   \\ 
 ESO 3020120 & 22.74$\pm$4.70 & 8.86$\pm$0.60 & 15.73 & 121.50 & 0.26 & 0.10 & 0.92 & 0.5046 \\ 
ESO 3050090 & 21.50$\pm$6.61 & 4.81$\pm$0.99 & 23.62 & 62.43 & 0.24 & 0.03 & 0.10 & 1.000 \\ 
 ESO 4250180 & 13.53$\pm$4.44 & 15.63$\pm$4.21 & 10.77 & 127.58 & 0.15 & 0.65 & 0.33 & 0.9237 \\ 
ESO 4880049 & 54.29$\pm$9.05 & 5.36$\pm$0.45 & 21.98 & 175.72 & 0.61 & 0.09 & 0.80 & 0.6176 \\ 
 F730 V1 & 41.37$\pm$4.02 & 10.31$\pm$0.41 & 14.22 & 257.24 & 0.47 & 0.47 & 11.62 & 0.  \\ 
 UGC 4115 & 142.22$\pm$45.79 & 1.53$\pm$0.82 & 50.77 & 131.09 & 1.55 & 0.004  & 0.007   & 1.000 \\ 
 UGC 11454 & 41.69$\pm$2.22 & 10.43$\pm$0.24 & 14.11 & 262.25 & 0.47 & 0.51 & 12.91 & 0.  \\ 
 UGC 11557 & 14.18$\pm$4.10 & 9.39$\pm$4.12 & 15.13 & 80.34 & 0.16 & 0.09 & 0.08 & 0.9997 \\ 
 UGC 11583 & 89.60$\pm$28.62 & 1.51$\pm$0.29 & 51.11 & 81.77 & 0.97 & 0.004   & 0.07 & 0.9998 \\ 
 UGC 11616 & 58.13$\pm$3.76 & 7.83$\pm$0.24 & 17.08 & 274.51 & 0.66 & 0.23 & 10.09 & 0.  \\ 
 UGC 11648 & 25.90$\pm$1.31 & 11.56$\pm$0.27 & 13.17 & 180.59 & 0.29 & 0.47 & 15.63 & 0.   \\ 
 UGC 11748 & 228.42$\pm$5.52 & 7.67$\pm$0.01 & 17.32 & 1056.49 & 2.58 & 2.96 & 58.47 & 0.  \\ 
 UGC 11819 & 48.62$\pm$2.84 & 9.63$\pm$0.33 & 14.88 & 282.48 & 0.55 & 0.39 & 2.11 & 0.0109 \\ 
 
\bottomrule
\hline \hline
\end{tabular}
}
\end{center}
\caption{We show the fitting parameters, derived quantities (see text for details) and the $\chi^2_{red}$ and $Q$-value for galaxies without photometry, for the Thomas-Fermi model without cut. Column (1) is the name of the galaxy; columns (2) and (3) give the fitting parameters; columns (4)--(7) give derived quantities and columns (8) and (9) give the $\chi^2_{red}$ and $Q$-value, respectively.}
\label{tab:TF-zerodisk}
\end{table*}
\end{center}

\begin{center}
\begin{table*}
\ra{1.3}
\begin{center}
\resizebox{16cm}{!} {
\begin{tabular}{@{}l r r r r r r r r r r @ {}}\toprule
    \hline \hline
    \multicolumn{9}{c}{Galaxy sample} \\
    \multicolumn{9}{c}{Thomas-Fermi with Cut DM Model}\\
    \hline 
   \multicolumn{9}{c}{Best Fitting Parameters}\\
   \hline 
\multicolumn{1}{c}{Galaxy} &
\multicolumn{1}{c}{$\rho_{s}$ }&
\multicolumn{1}{c}{$R$ } &
\multicolumn{1}{c}{$m_{\psi}$} &
\multicolumn{1}{c}{$\mu_{DM}$} &
\multicolumn{1}{c}{$M_{DM}(300$ pc$)$}& 
\multicolumn{1}{c}{$M_{DM}(r_{\text{max}})$}& 
\multicolumn{1}{c}{$\chi^2_{\text{red}}$} &
\multicolumn{1}{c}{$Q$-value}\\
\multicolumn{1}{c}{ }& 
\multicolumn{1}{c}{($10^{-3}M_\odot/$pc$^3$)}&
\multicolumn{1}{c}{(kpc) }& 
\multicolumn{1}{c}{(meV)}& 
\multicolumn{1}{c}{($M_\odot/$pc$^2$) }& 
\multicolumn{1}{c}{($10^{7}M_\odot$)}& 
\multicolumn{1}{c}{($10^{11}M_\odot$)}& 
\multicolumn{1}{c}{ }\\
\multicolumn{1}{c}{(1)}& 
\multicolumn{1}{c}{(2)}& 
\multicolumn{1}{c}{(3)}& 
\multicolumn{1}{c}{(4)}& 
\multicolumn{1}{c}{(5)}& 
\multicolumn{1}{c}{(6)}& 
\multicolumn{1}{c}{(7)}& 
\multicolumn{1}{c}{(8)}& 
\multicolumn{1}{c}{(9)}\\
 
\cmidrule[0.4pt](r{0.25em}){1-1}
\cmidrule[0.4pt](lr{0.25em}){2-2}
\cmidrule[0.4pt](lr{0.25em}){3-3}
\cmidrule[0.4pt](lr{0.25em}){4-4}
\cmidrule[0.4pt](lr{0.25em}){5-5}
\cmidrule[0.4pt](lr{0.25em}){6-6}
\cmidrule[0.4pt](lr{0.25em}){7-7}
\cmidrule[0.4pt](lr{0.25em}){8-8}
\cmidrule[0.4pt](lr{0.25em}){9-9}

 ESO 0140040 & 64.38$\pm$4.22 & 16.40$\pm$0.18 & 10.43 & 637.04 & 0.73 & 3.62 & 3.87 & 0.0007 \\ 
 ESO 0840411 & 4.69$\pm$1.35 & 11.82$\pm$3.61 & 12.98 & 33.43 & 0.05 & 0.08 & 0.10 & 0.9982 \\ 
ESO 1200211 & 19.57$\pm$4.08 & 2.29$\pm$0.06 & 38.77 & 27.03 & 0.22 & 0.003  & 0.29 & 0.9913 \\ 
 ESO 1870510 & 32.82$\pm$7.28 & 2.91$\pm$0.37 & 33.01 & 57.70 & 0.37 & 0.01 & 0.21 & 0.9935 \\ 
 ESO 2060140 & 63.56$\pm$3.97 & 6.78$\pm$0.04 & 18.80 & 259.98 & 0.72 & 0.25 & 3.83 & $3.3\times 10^{-6}$  \\ 
ESO 3020120 & 29.46$\pm$5.25 & 7.24$\pm$0.16 & 18.00 & 128.65 & 0.33 & 0.14 & 0.42 & 0.9245 \\ 
 ESO 3050090 & 21.50$\pm$6.61 & 4.81$\pm$0.99 & 23.62 & 62.43 & 0.24 & 0.03 & 0.10 & 1.000 \\ 
 ESO 4250180 & 13.53$\pm$4.44 & 15.63$\pm$4.21 & 10.77 & 127.58 & 0.15 & 0.65 & 0.33 & 0.9237 \\ 
 ESO 4880049 & 54.90$\pm$8.95 & 5.33$\pm$0.43 & 22.08 & 176.44 & 0.62 & 0.11 & 0.78 & 0.6356 \\ 
 F730 V1 & 66.70$\pm$6.19 & 7.58$\pm$0.1 & 17.45 & 305.14 & 0.75 & 0.37 & 9.01 & $7.2\times 10^{-10}$  \\ 
 UGC 4115 & 142.22$\pm$45.79 & 1.53$\pm$0.82 & 50.77 & 131.09 & 1.55 & 0.004  & 0.007   & 1.000 \\ 
 UGC 11454 & 47.99$\pm$2.00 & 9.47$\pm$0.12 & 15.05 & 274.12 & 0.54 & 0.52 & 10.91 & 0.   \\ 
 UGC 11557 & 14.18$\pm$4.10 & 9.39$\pm$4.12 & 15.13 & 80.34 & 0.16 & 0.09 & 0.08 & 0.9997 \\ 
 UGC 11583 & 89.60$\pm$28.62 & 1.51$\pm$0.29 & 51.11 & 81.77 & 0.97 & 0.004  & 0.07 & 0.9998 \\ 
 UGC 11616 & 86.20$\pm$4.68 & 6.13$\pm$0.07 & 20.11 & 318.68 & 0.97 & 0.25 & 5.29 & $2.1\times 10^{-8}$  \\ 
 UGC 11648 & 28.09$\pm$1.22 & 10.96$\pm$0.18 & 13.65 & 185.74 & 0.32 & 0.47 & 15.03 & 0.  \\ 
 UGC 11748 & 264.30$\pm$6.15 & 7.69$\pm$0.01 & 17.29 & 1225.92 & 2.98 & 1.53 & 24.22 & 0.   \\ 
 UGC 11819 & 52.90$\pm$2.62 & 8.98$\pm$0.19 & 15.59 & 286.44 & 0.60 & 0.49 & 1.19 & 0.2789 \\ 
 
\bottomrule
\hline \hline
\end{tabular}
}
\end{center}
\caption{We show the fitting parameters, derived quantities, the $\chi^2_{red}$ and the $Q$-value for the LSB galaxies for the Thomas-Fermi model with cut. The meaning of columns is the same as in Table \ref{tab:TF-zerodisk}.}
\label{tab:TF-zerodisk-cut}
\end{table*}
\end{center}

\begin{center}
\begin{table*}
\ra{1.3}
\begin{center}
\resizebox{16cm}{!} {
\begin{tabular}{@{}l r r r r r r r r r r @ {}}\toprule
    \hline \hline
    \multicolumn{9}{c}{Galaxy sample} \\
    \multicolumn{9}{c}{Flat Oscillaton DM Model}\\
    \hline 
   \multicolumn{9}{c}{Best Fitting Parameters}\\
   \hline 
\multicolumn{1}{c}{Galaxy} &
\multicolumn{1}{c}{$\rho_{s}$ }&
\multicolumn{1}{c}{$r_{s}$ } &
\multicolumn{1}{c}{$m_{\varphi}$} &
\multicolumn{1}{c}{$\mu_{DM}$} &
\multicolumn{1}{c}{$M_{DM}(300$ pc$)$}& 
\multicolumn{1}{c}{$M_{DM}(r_{\text{max}})$}& 
\multicolumn{1}{c}{$\chi^2_{\text{red}}$} &
\multicolumn{1}{c}{$Q$-value}\\
\multicolumn{1}{c}{ }& 
\multicolumn{1}{c}{($10^{-3}M_\odot/$pc$^3$)}&
\multicolumn{1}{c}{(kpc) }& 
\multicolumn{1}{c}{($10^{-27}$ eV)}& 
\multicolumn{1}{c}{($M_\odot/$pc$^2$) }& 
\multicolumn{1}{c}{($10^{7}M_\odot$)}& 
\multicolumn{1}{c}{($10^{11}M_\odot$)}& 
\multicolumn{1}{c}{ }\\
\multicolumn{1}{c}{(1)}& 
\multicolumn{1}{c}{(2)}& 
\multicolumn{1}{c}{(3)}& 
\multicolumn{1}{c}{(4)}& 
\multicolumn{1}{c}{(5)}& 
\multicolumn{1}{c}{(6)}& 
\multicolumn{1}{c}{(7)}& 
\multicolumn{1}{c}{(8)}& 
\multicolumn{1}{c}{(9)}\\
 
\cmidrule[0.4pt](r{0.25em}){1-1}
\cmidrule[0.4pt](lr{0.25em}){2-2}
\cmidrule[0.4pt](lr{0.25em}){3-3}
\cmidrule[0.4pt](lr{0.25em}){4-4}
\cmidrule[0.4pt](lr{0.25em}){5-5}
\cmidrule[0.4pt](lr{0.25em}){6-6}
\cmidrule[0.4pt](lr{0.25em}){7-7}
\cmidrule[0.4pt](lr{0.25em}){8-8}
\cmidrule[0.4pt](lr{0.25em}){9-9}

(1) ESO 0140040 & 68.95$\pm$6.67 & 6.39$\pm$0.35 & 1.00 & 440.40 & 0.78 & 5.28 & 3.14 & 0.005  \\ 
(2) ESO 0840411 & 4.78$\pm$1.47 & 5.03$\pm$1.70 & 1.27 & 24.05 & 0.05 & 0.08 & 0.09 & 1.0 \\ 
(3) ESO 1200211 & 33.68$\pm$26.22 & 0.70$\pm$0.28 & 9.10 & 23.70 & 0.37 & 0.004   & 0.17 & 1.0 \\ 
(4) ESO 1870510 & 35.31$\pm$9.45 & 1.16$\pm$0.21 & 5.52 & 40.96 & 0.39 & 0.01 & 0.15 & 1.0 \\ 
(5) ESO 2060140 & 87.65$\pm$18.27 & 2.33$\pm$0.25 & 2.75 & 204.29 & 0.99 & 0.39 & 1.96 & 0.02 \\ 
(6) ESO 3020120 & 34.19$\pm$14.29 & 2.68$\pm$0.79 & 2.39 & 91.60 & 0.39 & 0.16 & 0.25 & 0.99 \\ 
(7) ESO 3050090 & 22.41$\pm$7.53 & 1.98$\pm$0.49 & 3.23 & 44.45 & 0.25 & 0.03 & 0.09 & 1.0 \\ 
(8) ESO 4250180 & 14.90$\pm$6.45 & 6.19$\pm$2.15 & 1.03 & 92.27 & 0.17 & 0.65 & 0.28 & 0.95 \\ 
(9) ESO 4880049 & 61.12$\pm$12.50 & 2.09$\pm$0.24 & 3.07 & 127.48 & 0.69 & 0.11 & 0.56 & 0.83 \\ 
(10) F730 V1 & 126.25$\pm$22.69 & 2.20$\pm$0.19 & 2.91 & 277.74 & 1.42 & 0.51 & 3.03 & 0.006   \\ 
(11) UGC 4115 & 143.25$\pm$48.08 & 0.66$\pm$0.38 & 9.63 & 95.15 & 1.56 & 0.004  & 0.007  & 1.0 \\ 
(12) UGC 11454 & 53.91$\pm$4.61 & 3.68$\pm$0.17 & 1.74 & 198.11 & 0.61 & 0.53 & 8.48 & 0.  \\ 
(13) UGC 11557 & 14.36$\pm$4.39 & 4.05$\pm$1.91 & 1.58 & 58.10 & 0.16 & 0.09 & 0.07 & 1.0 \\ 
(14) UGC 11583 & 95.54$\pm$35.59 & 0.61$\pm$0.15 & 10.45 & 58.49 & 1.03 & 0.004  & 0.06 & 1.0 \\ 
(15) UGC 11616 & 101.60$\pm$13.32 & 2.31$\pm$0.16 & 2.77 & 234.48 & 1.15 & 0.30 & 3.46 & 0.0001   \\ 
(16) UGC 11648 & 31.12$\pm$2.06 & 4.27$\pm$0.16 & 1.50 & 132.94 & 0.35 & 0.48 & 13.14 & 0.   \\ 
(17) UGC 11748 & 514.54$\pm$55.02 & 1.97$\pm$0.11 & 3.25 & 1012.67 & 5.79 & 2.60 & 2.19 & 0.004  \\ 
(18) UGC 11819 & 56.70$\pm$4.49 & 3.59$\pm$0.19 & 1.78 & 203.35 & 0.64 & 0.52 & 0.76 & 0.71 \\ 

\bottomrule
\hline 
\hline
\end{tabular}
}
\end{center}
\caption{In this table we show the fitting parameters, derived quantities, the $\chi^2_{red}$ and the $Q$-value for the analysed LSB galaxies for the flat oscillaton DM model.}
\label{tab:SFDM-zerodisk}
\end{table*}
\end{center}


\section{Conclusions}\label{sec:Conclusions}

In this paper we have analysed three scalar field DM models using a RC LSB galaxies catalog reported in \cite{deBlok:2001mf}.
We found a better $\chi^2_{red}$ and $Q$-values when using the  SFDM model versus TF models.
TF approximation without cut presents problems to fit the RC because the profile density function, $\sin{kr}/kr$, has a strong decay when $R$ has the smallest values, giving negative values for the mass density, leading to instabilities of the circular orbits. In general TF approximation with and without a cut has problems to fit the observational data. 

For each model and for each galaxy in the catalog we computed two important quantities, the surface density $\mu_{DM}$ and the mass within 300 pc, $M_{DM}(300$ pc$)$ for the DM haloes. The values we obtained are roughly constant and independent of the absolute magnitude of the galaxies as was also found in Ref. \cite{2015arXiv151106740G}.
The mean value for $M_{DM}(300$ pc$)$ is consistent with the common mass for dwarf spheroidal galaxies of the order of $10^7$ $M_\odot$ reported in \cite{2008Natur.454.1096S} and independent of the DM model. On the contrary, the mean values of $\mu_{DM}$ do depend of the DM model, $225.7$ and $244.4$ $M_\odot/$pc$^2$ for TF models versus $186.7$ $M_\odot/$pc$^2$ for the flat oscillaton.




\end{document}